\documentclass[journal,11pt,draftclsnofoot,onecolumn]{IEEEtran}

\ifCLASSINFOpdf
\else
\fi
\usepackage{cite}
\usepackage{amsmath,amssymb,amsfonts}
\usepackage{amsthm}
\usepackage{algorithmic}
\usepackage{graphicx}
\usepackage{textcomp}
\usepackage{xcolor}
\usepackage{caption}
\newtheorem{theorem}{Theorem}
\newtheorem{lemma}{Lemma}
\newtheorem{definition}{Definition}
\newtheorem{remark}{Remark}

\newtheorem{cor}{Corollary}

\newtheorem{example}{Example}

\usepackage{epstopdf}
\usepackage{epsfig}
\usepackage{subfigure} %

\usepackage{color}

\begin{document}
%
\title{Multi-answer Constrained Optimal Querying: Maximum Information Gain Coding}

%
%
%

\author{\IEEEauthorblockN{
		Zhefan Li, Pingyi Fan$^*$, \IEEEmembership{Senior Member IEEE}\\}
	\thanks{This work was supported by the National Key R \& D Program of China under Grant 2021YFA1000500(4).}%
	\thanks{Z. Li and P. Fan are with the Department of Electronic Engineering, Tsinghua University, Beijing 100084, China and Beijing National Research Center for Information Science and Technology (BNRist), Beijing 100084, China (e-mail: lzf20@mails.tsinghua.edu.cn; fpy@tsinghua.edu.cn).}
}

\maketitle

\begin{abstract}
As the rapidly developments of artificial intelligence and machine Learning, behavior tree design in multiagent system or AI game become more important. The behavior tree design problem is highly related to the source coding in information theory. “Twenty Questions” problem is a typical example for the behavior tree design, usually used to explain the source coding application in information theory and can be solved by Huffman coding. In some realistic scenarios, there are some constraints on the asked questions. However, for general question set, ﬁnding the minimum expected querying length is an open problem, belongs to NP-hard. Recently, a new coding scheme has been proposed to provide a near optimal solution for binary cases with some constraints, named greedy binary separation coding (GBSC). In this work, we shall generalize it to D-ary cases and propose maximum information gain coding (MIGC) approach to solve the multi-answer decision constrained querying problem. The optimality of the proposed MIGC  is discussed in theory. Later on, we also apply MIGC to discuss three practical scenarios and showcase that MIGC has better performance than GBSC and Shannon Coding in terms of bits persymbol.  
\end{abstract}


%
\IEEEpeerreviewmaketitle

\section{Introduction}
%
%
%
%
\IEEEPARstart{C}{onstucting} a decision tree for identification is a fundamental problem in information theory and discrete mathematics \cite{r1}. As the rapidly developments of artificial intelligence and machine Learning, behavior tree design in multiagent system or AI game become more important. The behavior tree design problem is highly related to the source coding in information theory. “Twenty Questions” problem is a typical example for the behavior tree design, usually used to explain the source coding application in information theory and can be solved by Huffman coding. For the multiple agent systems, reinforcement learning\cite{b2}, Q-learning\cite{b3,b4}, and federated learning \cite{b5,b6,b7} may be the most important tools or frameworks to enhance the decision tree or behavior tree design. 

Let us review the well-known game “Twenty Questions”, where a player is required to determine the identity of an item in a certain set by asking a minimum number of “yes/no” questions \cite{r2}. It is equivalent to the binary source coding in information theory. Its optimal solution can be given by Huffman coding for the scenario without any constraint \cite{r3,r15}. 

In general, numerous variants of the game could be designed by adding constraints. One example is to restrict the times of the question query \cite{r4,r5,r14}, which can be solved by bounded-length Huffman coding. In the presence of large sets or multiple sources, the limited memory becomes the key constraint factor \cite{r6}, which may cause conventional source coding methods can not give a reasonable solution. In particular, if each question is endowed with a cost, the problem can be turned from minimum length into minimum cost \cite{r1}.  

 In this paper, we focus on the decision constrained case, where the question or decision set is limited. That is, one cannot divide the subtrees arbitrarily when building the decision tree. Generally, finding the optimal solution to it is an NP-hard problem\cite{r7}. Hence, it is common to utilize greedy algorithms to find some near optimal solutions rather than the exact solution \cite{r2}. In the literature, there exist heuristic algorithms that admit
 $O($log$n)$-approximation \cite{r8,r9}. Furthermore, if the decision constraints can be described by a graph, an approximation with $O($loglog$n)$ from the optimal one was given\cite{r1}. Some algorithms based on active learning were also presented in\cite{r10,r11} . 
 
 Recently, the authors in \cite{r2} proposed a Greedy Binary Separation Coding (GBSC) algorithm to the problem based on information theory. However, it is only suitable to the "yes/no" binary cases. Inspired by that, we generalize it to D-ary cases and propose a new algorithm MIGC referred to as Maximum Information Gain Coding. More specifically, the contribution can be mainly expressed as follows, 
 
 (i) We prove that for general distribution, the information quantity of the query is the entropy of the "answer". 
 
 (ii) We propose a Maximum Information Gain Coding algorithm to solve multi-answer decision constrained querying problem. 
 
 (iii) The optimality of MIGC is discussed in theory and some applications are given.
 
The rest of this paper is organized as follows. In Section II, we give a brief review of the GBSC algorithm. In Section III, we first build a generalized model for the multi-answer constrained problem and present the algorithm, Maximum Information Gain Coding (MIGC), to it. Then, we prove MIGC is better than Shannon Coding in terms of the average bits per symbol. In Section IV, three specific tasks are discussed to showcase the advantage of our proposed MIGC algorithm. Finally, the conclusion is given in Section V.


\section{Related Works}

\subsection{Binary Identification Problem}

First, we consider the binary case. Given a ramdom variable $X=\{1,2,...,N\}$ with distribution $P=\{p_1,p_2,\dots,p_N\}$, and a set of questions $A=\{S_1,S_2,...,S_K\}$, our objective is to find the minimum average query times by asking questions and present the selection of the $X$ in the sequel. Here, it is assumed that $S_i$ is the subset of $\emph{X}$, and the query can be designed as "Is $X$ in the set $S_i$ ?", where $S_i\in A$. If the query set is $2^{\emph{X}}$, then the optimal sulotion can be given by Huffman coding. However, for some complex structures, Huffman coding algorithm may not be workable. Let us see some examples.

\begin{example}
	Suppose $X\in\{1,2,3,4\}$, distribution of $X$ is $P=\{0.1,0.4,0.2,0.3\}$, and the query set is $A=\{\{1,2\},\{2,3\},\{3,4\}\}$. 
\end{example}

\begin{figure}[htbp]
	\centering
	\includegraphics[scale=0.33]{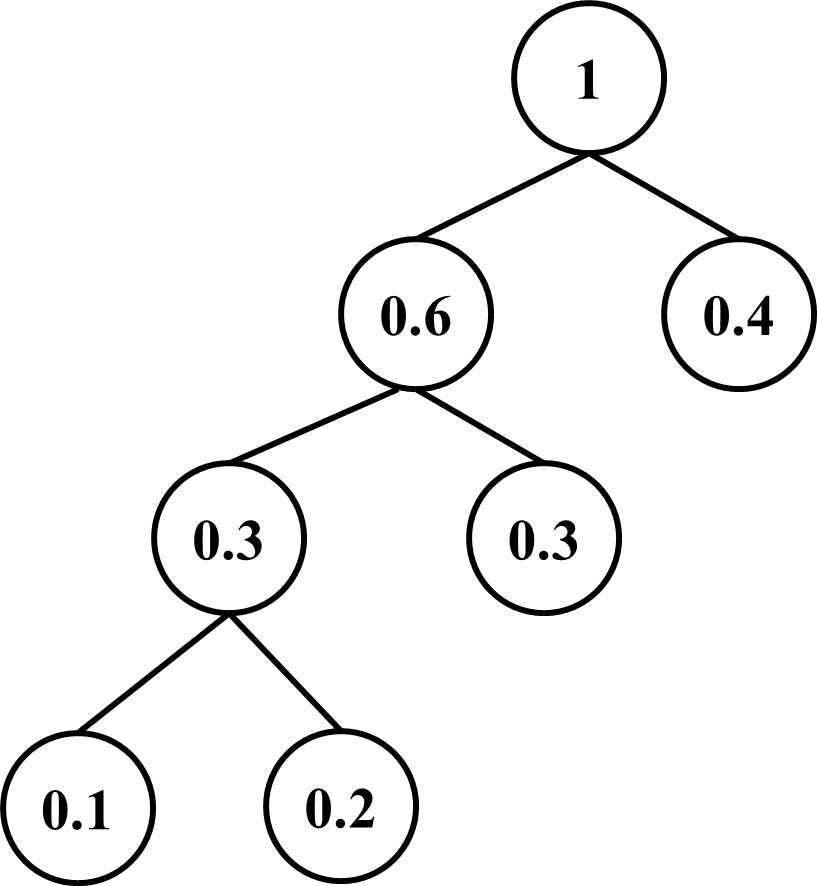}\\
	\caption{Huffman decision tree of the Example 1}
	\label{fig_1}
\end{figure}

The solution of Example 1 is shown as Huffman decision tree mode in Fig.\ref{fig_1}. From the Huffman decision tree, one can find that the query set should contain $\{1,3,4\}$ and $\{2\}$, but the query set in Example 1 did not contain them. That means the solution given by the Huffman decision tree is not valid to the problem in Example 1. 

\subsection{Greedy Binary Separation Coding}
To solve such a binary identification problem, recently, authors in \cite{r2} proposed an approximation algorithm based on a coding sheme, referred to as Greedy Binary Separation Coding (GBSC). The method constructs the decision tree from top to bottom and at each node, it chooses the query making the total probabilities of the left and right subtrees as close as possible. In this way, one can build the decision tree of Example 1 as in Fig.2    

\begin{figure}[htbp]
	\centering
	\includegraphics[scale=0.33]{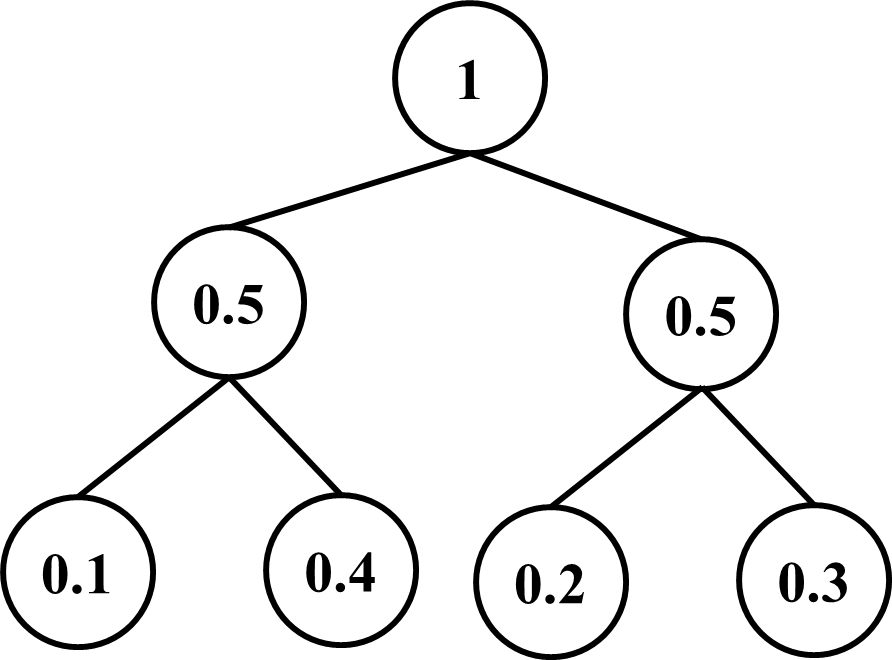}\\
	\caption{GBSC decision tree of the example 1}
	\label{fig_2}
\end{figure}

The intuition of GBSC is to greedily maximize the information gain at each node. It is proved that the mutual information between the target $X$ and the query $Q$ is equal to the binary entropy of the answer to the query. This indicates that it obeys the rule of the probability of answer "yes" to be close to $1/2$. However, in \cite{r2}, it is only proved GBSC is valid when $X$ is uniformly distributed. Now we will extend it and prove that for general discrete distributions, GBSC is still valid by using the near equal probability partition rule.  

Note that GBSC is only suitable to the query problem in binary cases. Inspired by maximizing the information gain of GBSC, we generalize it and propose the Maximum Information Gain Coding (MIGC) method to multi-answer cases .

\section{Maximum Information Gain Coding}
In this Section, we first introduce multi-answer constrained query problem and then present a new solution referred to as Maximum Information Gain Coding. Meanwhile we also analyze its optimality.

\subsection{Multi-answer Constrained Query Problem}
Different from the constrained query problem in binary cases, the query now has $D$ possible answers. Each answer can be uniquely mapped to one of $D$ disjoint subsets of $\mathcal{X}$. For simplicity, we first introduce some definitions. 

\begin{definition}
	The i-th query is mapped to $S_i=\{U_{i1},U_{i2},\dots,U_{iD}\}$, where $U_{ij}\subseteq \mathcal{X}$, $U_{ij} \cap U_{iq} = \emptyset, 1\le j<q\le D$ and $\emptyset$ represents the empty set.
\end{definition} 
 That is, the i-th query can be described as "Which subset $U_{ij}$ is $X$ in", where $U_{ij}\in S_i$. 
 
We now formally desciribe MIGC.
\begin{definition}
	$S=\{U_1,U_2,...,U_D\}$ is a $D$-ary partition of $\mathcal{X}$, if $\bigcup_{1\le j\le D}U_j =X$ and $U_j\cap U_q = \emptyset, \forall 1\le j<q\le D$ 
\end{definition} 

\begin{definition}
	For a subset $U\subseteq X$, its total probability
	\begin{equation}
		p(U)=\sum_{x_i\in U} p_i
	\end{equation} 
\end{definition}

\begin{definition}
	The entropy of the partition $S=\{U_1,U_2,...,U_D\}$ is defined as
	\begin{equation}
		H(S)=-\sum_{j=1}^D p(U_j)logp(U_j)
	\end{equation}
\end{definition}

\begin{definition}
	A partition $S^*$ of $\mathcal{X}$ is optimal, if for any  partition $S$, $H(S^*)\ge H(S)$
\end{definition}

Similar to GBSC, the rule from top to bottom is still adopted for constructing the MIGC decision tree. The process can be described as follows. At the root, we choose the query that makes the partition of $\mathcal{X}$ optimal as defined in Definition 5, and split the tree
according to the partition. We repeat it recursively at each node, and choose the query corresponding to the optimal partition. In this way, it is easily to know that GBSC is the special case of MIGC for $D=2$.

\begin{example}
	Suppose $X\in\{1,2,3,4,5\}$, distribution of $X$ is $P=\{0.1,0.2,0.3,0.15,0.25\}$, $D=3$ and the query set is unconstrained. 
\end{example}

Based on the rule of MIGC, the corresponding decision tree of Example 2 is shown in Fig.3. It gives clearly solution to the query problem in Example 2.
\begin{figure}[htbp]
	\centering
	\includegraphics[scale=0.33]{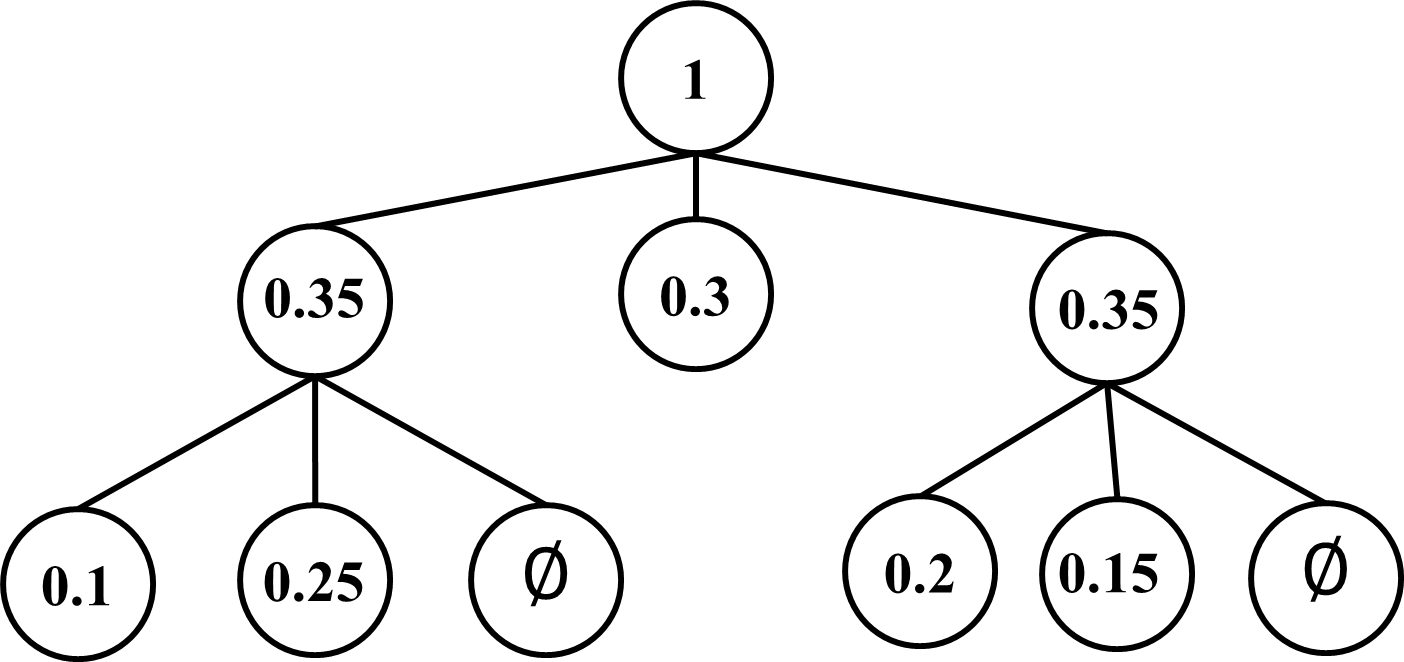}\\
	\caption{MIGC decision tree of the Example 2}
	\label{fig_3}
\end{figure}

\subsection{Intuition
}\label{AA}
In this subsection, we illustrate why the MIGC can approach maximum entropy at each query. For a query $Q$ on the node $X\in \mathcal{X}=\{x_1,x_2,\dots,x_N\}$, each $x_i$ gives a $D$-ary answer. The probability of getting answer $j$ is given by
\begin{equation}
	P(Q=j)=p_j=\sum_{x_i \in \mathcal{X} \land \,x_i\, gives\, answer\, j} p(x_i), 
\end{equation}
The information gain of query $Q$ with respect to node $X$ is 
\begin{equation}
	\begin{split}
		I(X;Q)&=H(X)-H(X|Q) \\
		&=H(X)-\sum_{j=1}^D p_jH(X|Q=j)
	\end{split}
\end{equation}

\begin{lemma}
	\begin{equation}
		I(X;Q)=H(Q)
	\end{equation}
\end{lemma}

In \cite{r2}, it proved $I(X;Q)=H(Q)$ is true when $X$ is discrete uniformly distributed. Now we will show that for any discrete distribution of $X$, Lemma 1 always holds.

\begin{proof}
\begin{equation}
	\begin{split}
		I(X;Q)&=H(X)-\sum_{j=1}^D p_jH(X|Q=j) \\
		&=H(X)+\sum_{j=1}^D p_j \sum_{n=1}^N \frac{P(x_i,Q=j)}{p_j}log\frac{P(x_i,Q=j)}{p_j} \\
		&=H(X)+\sum_{j=1}^D\sum_{n=1}^N [P(x_i,Q=j)logP(x_i,Q=j)\\
		&\quad-P(x_i,Q=j)logp_j]
	\end{split}
\end{equation}

Notice that 
\begin{equation}
	\sum_{j=1}^D\sum_{n=1}^N P(x_i,Q=j)logP(x_i,Q=j)=-H(X)
\end{equation}
\begin{equation}
	\sum_{j=1}^D\sum_{n=1}^N -P(x_i,Q=j)logp_j=-\sum_{j=1}^D p_jlogp_j =H(Q)
\end{equation}

Then we have 
\begin{eqnarray}
	I(X;Q)=H(X)-H(X)+H(Q)=H(Q) 
\end{eqnarray}
\end{proof}

Lemma 1 indicates the information we gain from the query about $X$ is equal to the entropy of the answer. In fact, the uniform distribution of the answer maximizes the information gain, which means each answer should appear with the same probability. 

Since different answers divide the $\mathcal{X}$ into non-overlap partitions, $H(Q)$ can also be seen as the partition entropy as defined in section IIIA.

\subsection{Near-optimality}
In this subsection, we discuss the optimality of MIGC. For simplicity, we only consider unconstrained condition. It is well known Huffman coding gives the optimal solution to the  unconstrained query problem. However, both GBSC and MIGC may not reach the optimal solution for the unconstrained problem. That means there exists a gap between Huffman coding approach and GBSC or MIGC in terms of bits persymbol. In fact, GBSC for unconstrained query problem looks like the Fano's source coding, but the main difference from Fano's source coding is that Fano requires the order ranking of the answer probability from the least to the largest. GBSC does not require such preprocessing. In \cite{r2}, it proves that GBSC is at least as good as Shannon coding\cite{r15} and better than Fano's approach in terms of the upper bound of the average coding length. It is remarkable that the coding length for each symbol in GBSC is no more than the corresponding coding length in Shannon coding. In the sequel, we will prove it for MIGC. 

\begin{theorem}
	Assume that the expected coding length is $L_m$ for MIGC, and $L_s$ for Shannon coding, then $L_m\le L_s$ 
\end{theorem}

\begin{cor}
Assume that the expected code length is $L_m$ for MIGC, then $L_m<H(X)+1$ 
\end{cor} 

Since $L_s<H(X)+1$, using Theorem 1, one can easily get the upper bound of $L_m$ in Corollary 1.

Now we shall prove Theorem 1, and we first prove the Lemma 2. 

\begin{lemma}
	Assume the code length for symbol with probability $p$ is $L_D$ for MIGC, then $L_D\le\lceil logp \rceil$
\end{lemma}

Note that this is a much stronger result in source coding theory. For the binary case, it has been proven in \cite{r2} by contradiction. However, such method cannot be directly generalized to D-ary cases. This is because each node in a binary tree only has two children, which can be easy to discuss separately. But for D-ary scenario, it is necessary to sort the children nodes at each step. Now we give the detail of the proof as follows.

\begin{figure}[htbp]
	\centering
	\includegraphics[scale=0.4]{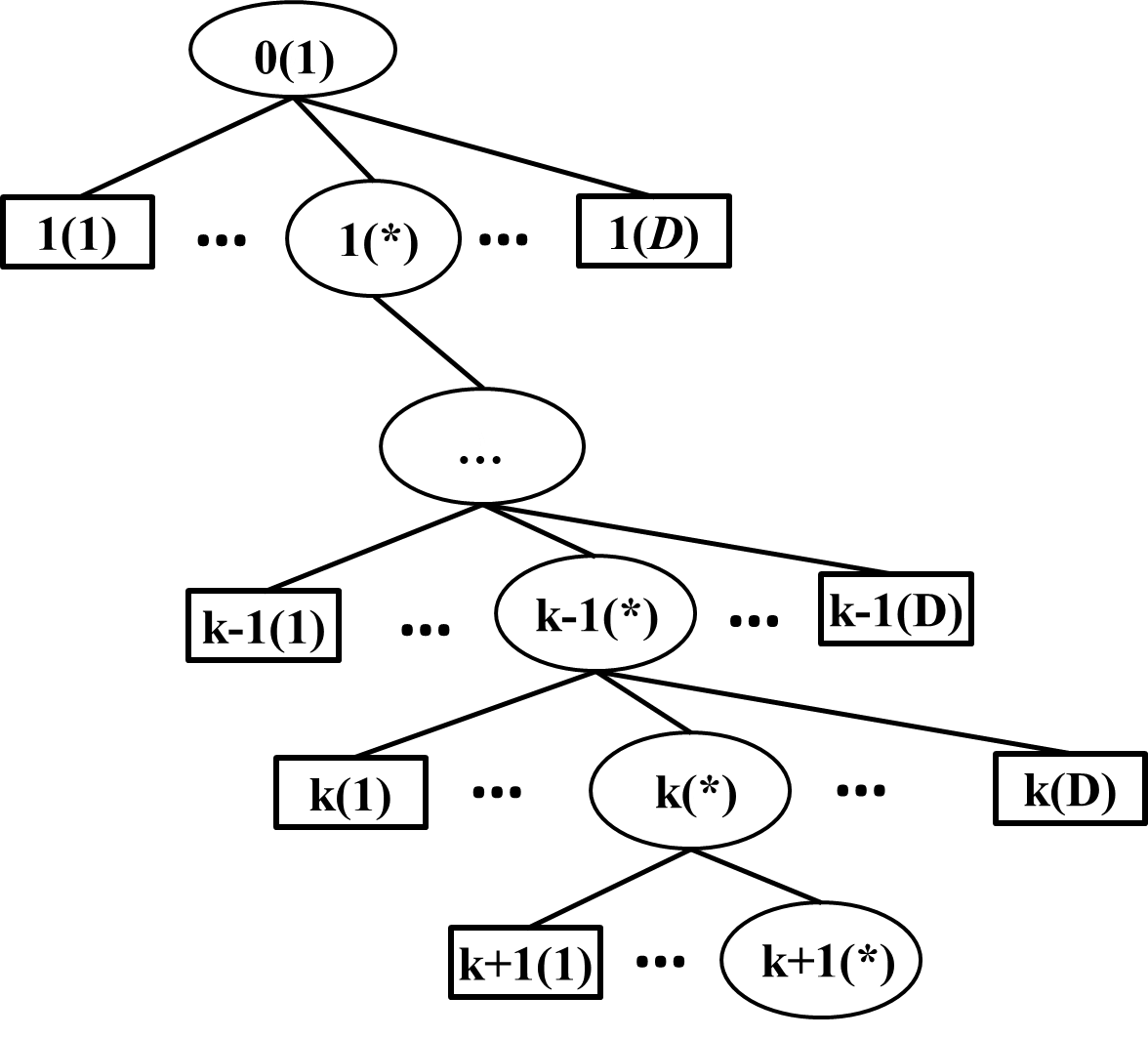}\\
	\caption{Illustration for the proof of Lemma 2.}
	\label{fig_proof}
\end{figure}
\begin{proof}
	On the children nodes sorting process, it is illustrated in Fig.\ref{fig_proof}, where a circle represents a single node while a rectangle represents a leaf or a subtree. The content in the nodes are indices of the nodes. $p_{n(i)}$ is the sum of probabilities of all the children nodes of node $n(i)$. 
	
	Without lose of generality, we assume for any $n$ and $i<j$, $p_{n(i)}<p_{n(j)} $. 
	
	Similar to \cite{r2}, we will prove it by contradiction. Without lose of generality, assume $p_{k+1(*)}\ge D^{-k}$ where $p_{k+1(*)}$ is the maximum among its sister nodes. For $n\le k$, $n(*)$ represents the nodes on the path from $k+1(*)$ to the root. 
	
	Then we want to prove the following two propositions hold for any integer $1\le n\le k$ by mathematical induction.
	
	(i) $p_{n(D)}\ge D^{-n}$
	
	(ii) $p_{n-1(*)}\ge D^{-(n-1)}+p_{k+1(1)}$
	
	\begin{remark}
		if (i) and (ii) holds, then $p_{0(1)}\ge 1+p_{k+1(1)}>1$ , causing a contradiction.
	\end{remark}

	\textbf{Basic step}.
	
	 First, we prove the case when $k=n$. Since 
	\begin{equation}
		\begin{split}
			&p_{k(D)}\ge p_{k(*)} \\
			&p_{k(*)}\ge p_{k+1(1)}+p_{k+1(*)},
		\end{split}
	\end{equation}
 then
 \begin{equation}
 	p_{k(D)}\ge p_{k+1(*)}\ge D^{-k},
 \end{equation}
  Proposition (i) is true. 
  
  If $p_{k(*)}-p_{k(1)}>p_{k+1(1)}$, then moving node $k+1(1)$ from $k(*)$ to $k(1)$ will make $|p_{k(*)}-p_{k(1)}|$ smaller. According to the convex property of entropy, this will increase the partition entropy of node $k-1(*)$, which contradicts to the optimal partition defined in Definition 5. Hence, 
  \begin{equation}
  	p_{k(*)}-p_{k(1)}\le p_{k+1(1)},
  \end{equation}
 so we have 
 \begin{equation}
 	p_{k(1)}\ge p_{k(*)}-p_{k+1(1)}\ge p_{k+1(*)}\ge D^{-k}.
 \end{equation}
 And
	\begin{equation}
		\begin{split}
			p_{k-1(*)}&=\sum_{i=1}^D p_{k(i)}\ge (D-1)p_{k(1)} +p_{k(*)} \\
			&\ge (D-1)D^{-k}+D^{-k}+p_{k+1(1)} \\
			&\ge D^{-(k-1)}+p_{k+1(1)}
		\end{split}
	\end{equation}
	Proposition (ii) is true.

	\textbf{Inductive step}.
	
	 Assume the propositions are true for $n=m+1(1\le m\le k-1)$. Then 
	\begin{equation}
		p_{m(D)}\ge p_{m(*)}\ge D^{-m}+p_{k+1(1)}\ge D^{-m},
	\end{equation}
	Proposition (i) is true. 
	
	If $p_{m(*)}-p_{m(1)}>p_{k+1(1)}$, then moving node $k+1(1)$ from $m(*)$ to $m(1)$ will make $|p_{m(*)}-p_{m(1)}|$ smaller. According to the convex property of entropy, this will increase the partition entropy of node $m-1(*)$,  which contradicts to the optimal partition defined in Definition 5. Hence,
	\begin{equation}
		p_{m(*)}-p_{m(1)}\le p_{k+1(1)},
	\end{equation}
 so we have 
 \begin{equation}
 	p_{m(1)}\ge p_{m(*)}-p_{k+1(1)}\ge D^{-m}.
 \end{equation} 
And  
	\begin{equation}
		\begin{split}
			p_{m-1(*)}&=\sum_{i=1}^D p_{m(i)}\ge (D-1)p_{m(1)} +p_{m(*)} \\
			&\ge (D-1)D^{-m}+D^{-m}+p_{k+1(1)} \\
			&\ge D^{-(m-1)}+p_{k+1(1)}
		\end{split}
	\end{equation}
	Proposition (ii) is true. This demonstrates when $n=m$, the two propositions also hold.   
	
	By using Remark 1 and the discussion above, Lemma 2 is proved.              

\end{proof} 

\section{Some Practical Scenario Discussion}
In this section, we use three  practical scenarios to compare the performance of MIGC and other coding algorithms.

\subsection{Average Coding Length of General Discrete Distributed X for D=3 Without Constraints}\label{section.accuracy_MI}
Set $D=3$ and $N\in [3,12]$, and we generate $t=10000$ samples of discrete distribution randomly for each $N$ where each sample represents a discrete probability distribution. Fig.5(a) shows the average coding length of three different algorithms changes with $N$. It is observed that when $N>5$, the  average coding length of MIGC is between the Huffman coding and Shannon coding. Note that the Huffman coding is the optimal.

\begin{figure}[htbp]
	\centering
	\includegraphics[scale=0.55]{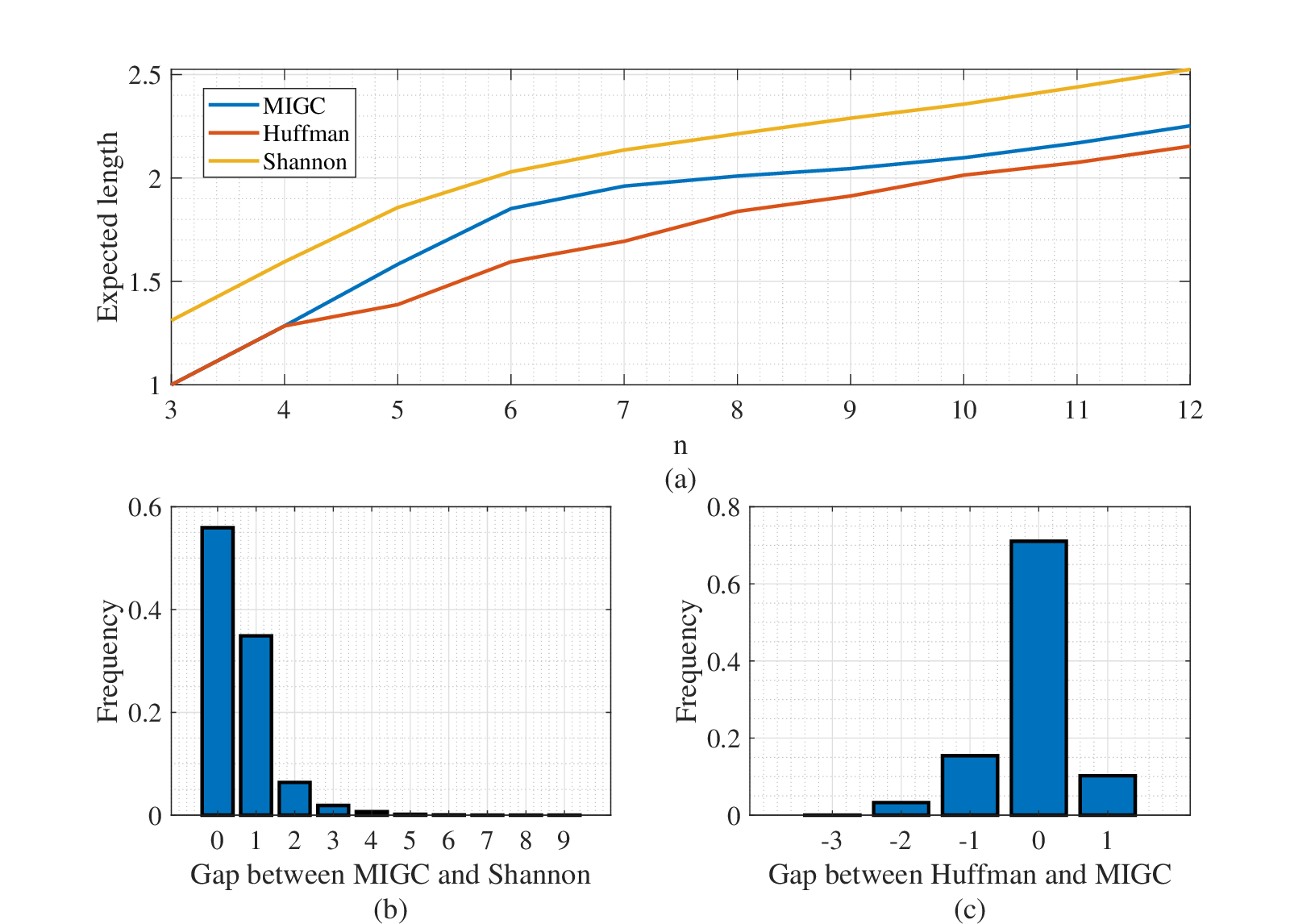}\\
	\caption{Performance of MIGC, Huffman coding and Shannon coding. (a) shows how the expected code length change with $N$. (b)shows the gap of code length between MIGC and Shannon coding when N=10. (c) shows the gap of code length between Huffman coding and MIGC when N=10.} 
	\label{fig_3}
\end{figure}

By setting $N=10$, Fig.5(b) shows the gap between the coding length of MIGC and that of Shannon coding for each symbol, it is easily seen that all the results are non-negative, which validates Lemma 2. Fig.5(c) shows the gap between the coding length of Huffman coding and that of MIGC for each symbol. Interestingly, although Huffman coding has less average coding length than MIGC, MIGC has more symbols that have less coding length than that of Huffman coding. And for a few symbols, their coding length of MIGC can be shorter than that of Huffman coding by 3.

\subsection{Application to DNA Detection}
\begin{figure}[htbp]
	\centering
	\includegraphics[scale=0.5]{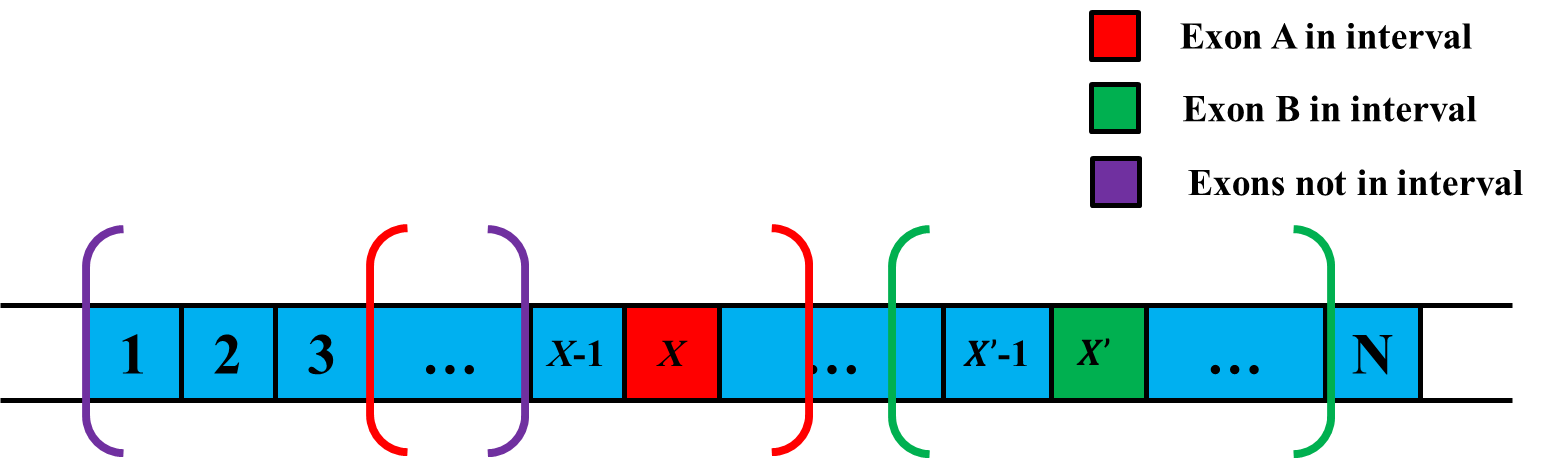}\\
	\caption{DNA detection} 
	\label{fig_DNA}
\end{figure}

The identification of genes is important in molecular biology, and is usually used to determine genomic sequences of different organisms\cite{r12}. An exon, which is an interval of the DNA and does not overlap with each other. A gene is a set of exons\cite{r13}. For simplicity, assume the target gene only contains one exon. Here, different from \cite{r2}, we set two target genes A and B for the detection work. Each time, one can detect a continuous interval of the DNA, and acquire four kinds of results, ``detect A", ``detect B", ``detect A and B", ``detect neither". Due to the position of the target genes being fixed, the DNA detection is to find the optimal detection querying with minimum expected detection times. The procedure is shown in Fig.\ref{fig_DNA}.

\begin{figure}[htbp]
	\centering
	\includegraphics[scale=0.4]{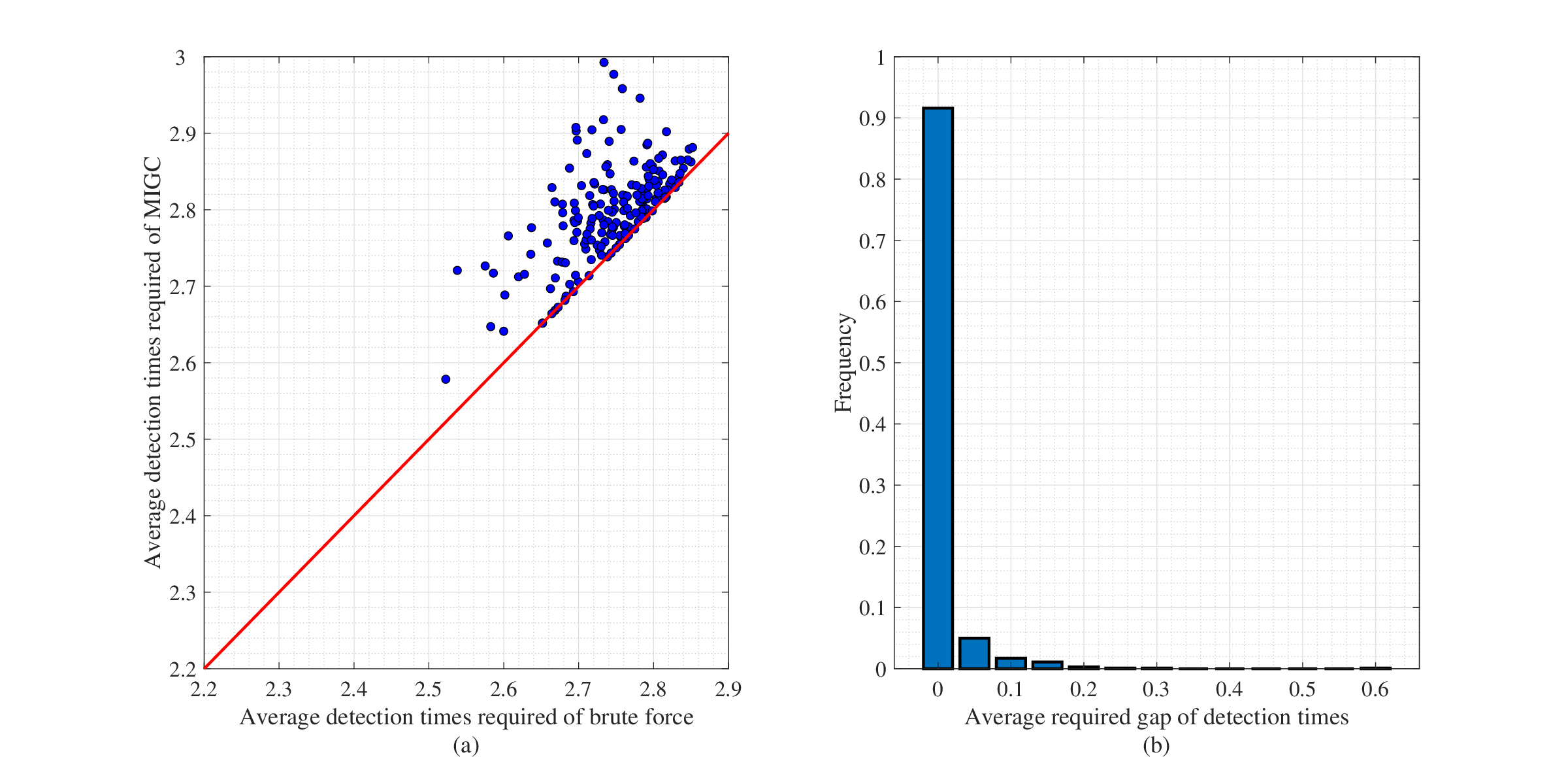}\\
	\caption{The DNA detection by using brute force and MIGC} 
	\label{DNA_BF_MIGC}
\end{figure}

\begin{figure}[htbp]
	\centering
	\includegraphics[scale=0.7]{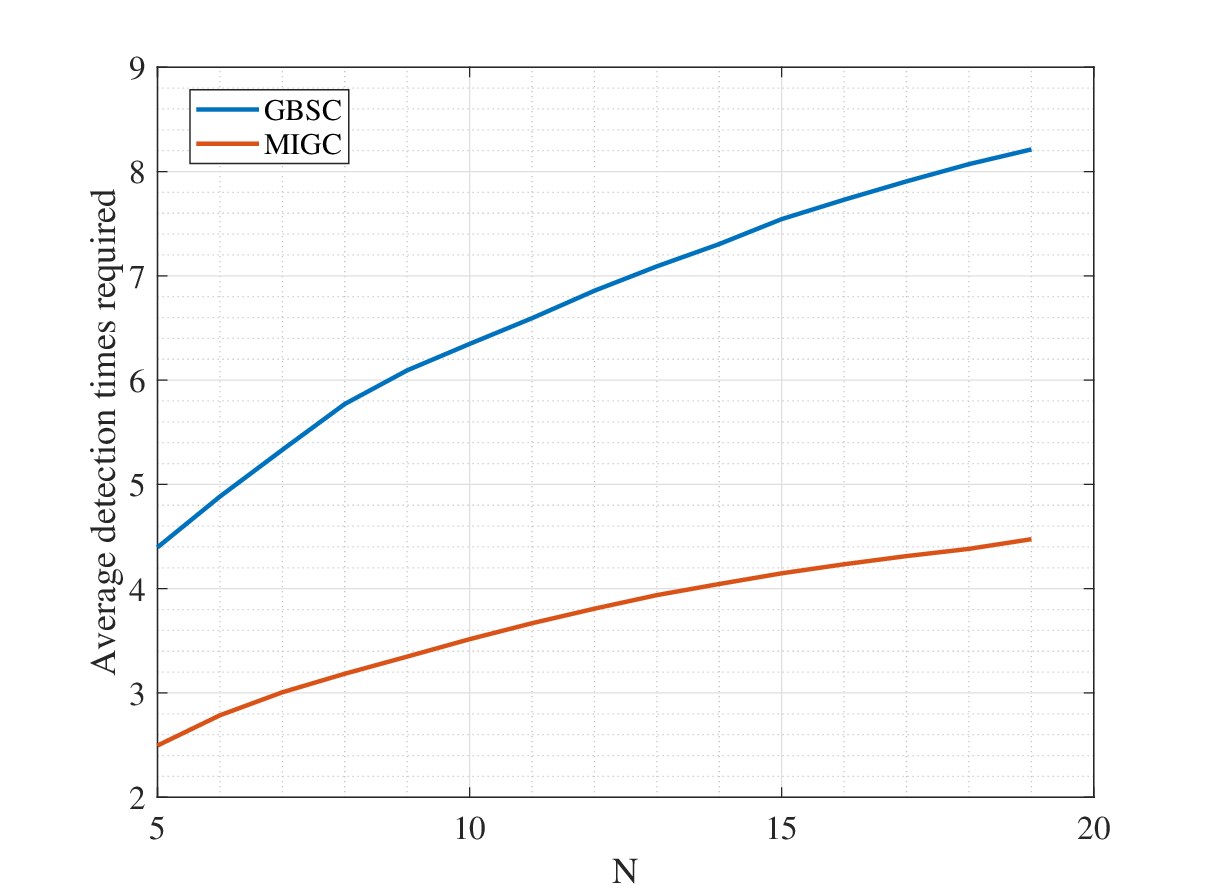}\\
	\caption{The DNA detection by using GBSC and MIGC} 
	\label{DNA_GBSC_MIGC}
\end{figure}

 DNA detection can be viewed as decision constrained coding problem. Set the number of exons $N=6$ and generate $t=10000$ samples of discrete distributions for testing. Fig.\ref{DNA_BF_MIGC} shows the gap between the MIGC and the optimal solution by using brute force. In Fig.\ref{DNA_BF_MIGC}(a), the red line represents the minimum average detection times, while the blue dots represent the results provided by MIGC. One can see that most blue dots are very close to the red line, which illustrates the effectiveness of MIGC. Fig.\ref{DNA_BF_MIGC}(b) shows the distribution of the gap between MIGC and brute force. It demonstrates that the gap is less than 0.16 detection times with 95$\%$ probability while the optimal average detection times is 2.16. Fig.\ref{DNA_GBSC_MIGC} compares the performance of MIGC and GBSC when $N$ changes, illustrating that MIGC has less average detection times than GBSC.

\subsection{Application to Battleship Games with 2 players}
Battleship is a popular strategy game\cite{r16}. In conventional battleship game, it requires two players to place "ships" on 10 by 10 boards, which are hidden to the opponent. The rules are expressed as follows, (i) the two players will play in turn to choose a grid $(i,j)$ (ii) if all the grids of one ship have been chosen by its owner's opponent, then the ship is "bombed" (iii) when one player's ships are all "bombed", the game ends and another player will win the game.

Now we modify the battleship games as follows.

Assume two players only place their ships on the same 10 by 10 board randomly. They do not participate in the following procedure of the game. A third player tries to bomb the ships placed by the former two players. Each time, the third player can get the result from $\{$``hit player1's ship", ``hit player2's ship", ``hit nothing"$\}$. The target for the third player is to sink all the ships with fewest tries (bombs). 

(i) If we do not distinguish the owner of the ships as in \cite{r2}, the problem reduces to the binary case, i.e., the bombing result for the third player is either "hit" or not. This can be well handled by GBSC. 

(ii) If we take the owner of the ships into account, it corresponds to the multi-answer constrained cases ( each time, the third player can get the result from $\{$``hit player1's ship", ``hit player2's ship", ``hit nothing"$\}$). In this case, GBSC does not work and we ultilize MIGC instead. The query set is the coordinates $\{(i,j),1\le i,j\le10\}$ of the board. For each try of the third player, one needs to maximize the entropy of the hitting result. 

Assume each of the former two players place a 1 by 5 and a 1 by 3 ships on the board, i.e., there are total 4 ships. We generate $n=3^{12}$ possible board layouts randomly, and the game starts by choosing one of them as target board (every legal board is equally possible). Since each query (bombing) acquires in average at most $log_2 3$ bits of information, the theoritical minimal number of average tries is no less than 12.

\begin{figure}[htbp]
	\centering
	\includegraphics[scale=0.6]{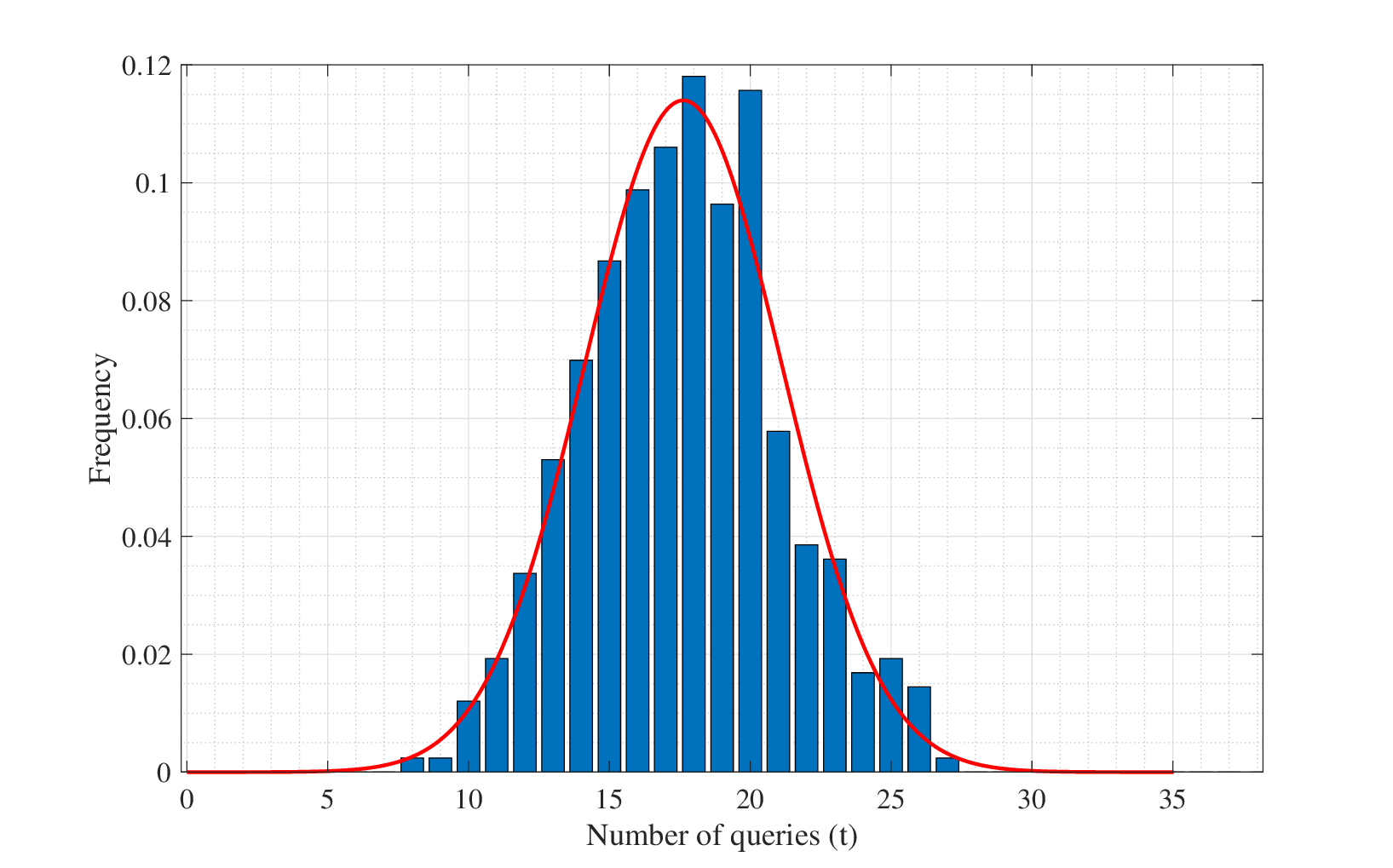}\\
	\caption{Distribution of the average try times for the battleship problem}
	\label{distribution_battleship}
\end{figure}

\begin{figure}[htbp]
	\centering
	\includegraphics[scale=0.8]{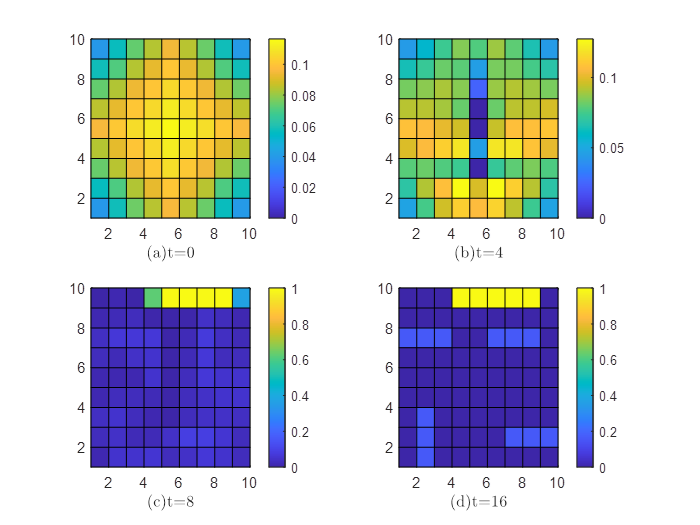}\\
	\caption{The process of playing battleship}
	\label{process}
\end{figure}

Fig.\ref{distribution_battleship} shows the distribution of the average try times of the third player. Fig. \ref{process} illustrates an example showcase of the game where color of each grid represents the probability of whether there exist ships of player 1. As the game progresses, the probability of the ships being bombed will become gradually larger. From Fig.\ref{process}(d), after trying bombs of $t=16$ times, one ship of size 1 by 5 can be bombed definitely next time and the location of the other ship of size 1 by 3 has been narrowed to four possibilities.

\begin{figure}[htbp]
	\centering
	\includegraphics[scale=0.5]{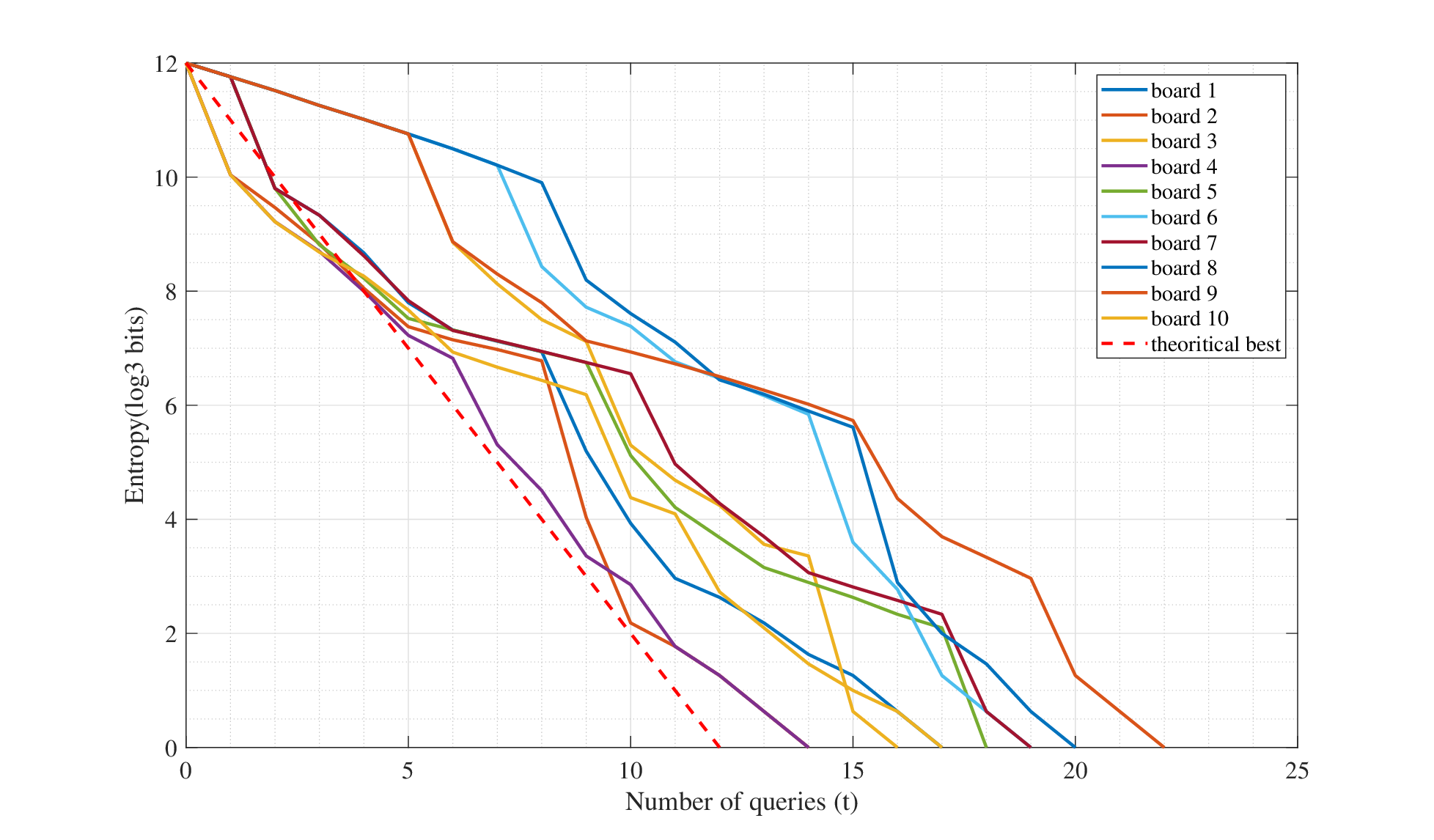}\\
	\caption{How the uncertainty of the board changes with the number of tries.}
	\label{info_change}
\end{figure}

Fig. \ref{info_change} shows how MIGC plays the game from the perspective of information theory. Since every target board has the same probability of being chosen, we can calculate the entropy of the board by $H(X_t)=log_3 |X_t|$, where $X_t$ represents the set of boards meeting the hitting results after $t$ tries, and $|\cdot|$ represents counting the number of elements in the set. We choose 10 boards randomly and show how the entropy or uncertainty of each board changes with the number of tries. The red dashed line represents the theoretical best result in average, which means each query (bombing) acquires $log_2 3$ bits of information. We can observe two types of the declination of entropy silimlar to the results in \cite{r2}. Most of times, the first few tries will not hit ships and the algorithm needs to explore the board, e.g., $t<5$ for board 6. At that stage, the declination of entropy is relatively slow. When a "hit" happens, the entropy drops rapidly until the information provided by this "hit" is fully ultilized. The process ends when the entropy reduces to the zero.

The modified battleship problem for the third player illustrates that MIGC can give an effective solution to the problem than that of GBSC.

\section{Conclusion}
In this paper, we discussed an open problem in information theory, the multi-answer constrained optimal query problem and proposed an Maximum Information Gain Coding, MIGC, solution to it. We also discussed its optimality of MIGC in theory. It is proved that MIGC has better performance than that of Shannon coding in terms of the average coding length. Furthermore, an interesting result is obtained that the coding length persymbol in MIGC is less than that by Shannon coding for $D\ge3$. Finally, we discuss three different scenarios and show that MIGC provides better results than GBSC and Shannon coding. It is expected the MIGC method can be combined with or embedded in some machine learning frameworks, i.e., federated learning mode in \cite{r20,r21} and got more applications in the future.

\section*{Acknowledgment}

The authors would like to appreciate the support of the National Key R \& D Program of China No. 2021YFA1000504. The authors thank the members of Wistlab of Tsinghua University for their good suggestions and discussions.

\appendices

\ifCLASSOPTIONcaptionsoff
  \newpage
\fi



%
\bibliographystyle{IEEEtran}
\bibliography{reference}
%







\end{document}